\documentclass[%
 reprint,
  superscriptaddress,
 amsmath,amssymb,amsfonts,
 aps,
 prl,
 floatfix,
 longbibliography
]{revtex4-2}

\usepackage[utf8]{inputenc}

\usepackage[pdftex]{graphicx} \graphicspath{{}}
\usepackage{float} \usepackage{color}

\usepackage[pdftex,colorlinks=true]{hyperref}
\hypersetup{
  colorlinks=true,
  linkcolor=blue,
  urlcolor=cyan,
}

\usepackage{mathtools}

\DeclarePairedDelimiterX\braket[2]{\langle}{\rangle}{#1 \delimsize\vert #2}
\DeclarePairedDelimiterX\expval[3]{\langle}{\rangle}{#1 \delimsize\vert #2  \delimsize\vert #3}
\DeclarePairedDelimiterX\proj[2]{\delimsize\vert#1\rangle}{\langle#2\delimsize\vert}{ }

\usepackage{siunitx}

\def\sectionn#1{\textit{#1:}}

\begin{document}

\title{
Tunable momentum pair creation of spin excitations in dipolar bilayers
}

\author{Thomas Bilitewski}
\affiliation{Department of Physics, Oklahoma State University, Stillwater, Oklahoma 74078, USA}
\author{G. A. Domínguez-Castro}
\affiliation{Institut f\"ur Theoretische Physik, Leibniz Universit\"at Hannover, Appelstr. 2, D-30167 Hanover, Germany}
\author{David Wellnitz}
\author{Ana Maria Rey}
\affiliation{JILA, National Institute of Standards and Technology and Department of Physics, University of Colorado, Boulder, CO, 80309, USA}
\affiliation{Center for Theory of Quantum Matter, University of Colorado, Boulder, CO, 80309, USA}
\author{Luis Santos}
\affiliation{Institut f\"ur Theoretische Physik, Leibniz Universit\"at Hannover, Appelstr. 2, D-30167 Hanover, Germany}

\date{\today}

\begin{abstract}
We study the temporal  growth and spatial propagation of quantum correlations in a two-dimensional bilayer realising a  spin-1/2 quantum  XXZ  model with couplings mediated by long-range and anisotropic dipolar interactions. Starting with an initial state consisting of  spins with opposite magnetization in each of the layers, we predict a dynamic instability that results, at short times, in the creation of correlated pairs of excitations at specific momenta at exponentially fast rates and entanglement between spatially separated modes. The momentum structure of the created pairs can be controlled via the dipolar orientation, the layer separation or the dipolar couplings. The predicted behavior remains  observable at very low filling fractions, making it accessible in state-of-the-art experiments with Rydberg atoms, magnetic atoms, and polar molecule arrays.
\end{abstract}

\maketitle
Anisotropic dipolar interactions  controllable via electromagnetic fields   offer unique opportunities for the implementation  of iconic models of quantum magnetism  relevant  for fundamental science and for the development of  novel quantum technologies.  In recent years,  great progress has been made on the implementation of  dipole-induced spin exchange interactions  in  fully controllable  quantum systems of  polar molecules \cite{Baranov_ChemicalReviews_112_2012,Bohn_Science_357_2017,moses2017new},  magnetic  atoms \cite{chomaz2022dipolar} and  Rydberg atoms \cite{browaeys2020many,RevModPhys.82.2313}. However,   most of the investigations so far have been  targeted to the single excitation regime \cite{Leseleuc2019} or to  the case of  multiple excitations   characterized via collective observables \cite{
Yan_Nature_501_2013, dePaz_PRL_111_2013, Hazzard2014, Lepoutre2019, Garbados2020,Patscheider_PRR_2_2020,Bilitewski_PRL_2021,Tobias_Science_375_2022,Bilitewski_dipolar_pair_2022,Li2023}.  Nevertheless, 
recent experimental developments on quantum gas microscopes \cite{christakis2022probing,NatPhysGross2021} and optical tweezers \cite{NatPhysKaufman2021,holland2022ondemand,bao2022dipolar,Anderegg2019,zhang2020forming} that allow for the spatial-resolved control of correlations  at the single particle level are opening a window  to explore  rich and intriguing  quantum phenomena enabled by dipolar spin models. 

In this work, we study the temporal and spatial growth of correlations during the many-body dynamics of an array of spin-1/2 frozen dipoles  confined in two separated two-dimensional layers (see Fig.~\ref{fig:model}(a)). This system, implementable  for example  using optical lattices or tweezer arrays,  realises a quantum XXZ spin model with dipolar couplings.  By preparing the two layers in opposite spin states, as in recent experiments on polar molecules~\cite{Tobias_Science_375_2022}, one  creates a dynamically unstable state from which correlated pairs of spin excitations  develop  and  grow at an exponential rate, at least at short times. These correlated pairs manifest in the spin structure factor, which develops intriguing momentum patterns  controllable by  both  the separation of the layers, and the magnitude and orientation of the dipole moments. 

The build up of spin correlations can be explained using a Bogoliubov analysis, which uncovers a dynamical instability in specific tunable momentum modes. We validate the Bogoliubov predictions of the pair creation patterns by numerical simulations of the full spin dynamics, and show that pattern formation remains robust even for very low lattice fillings, making it observable in state of the art experiments, without requiring unit-filling.  



\begin{figure}[t]
\includegraphics[width=\columnwidth]{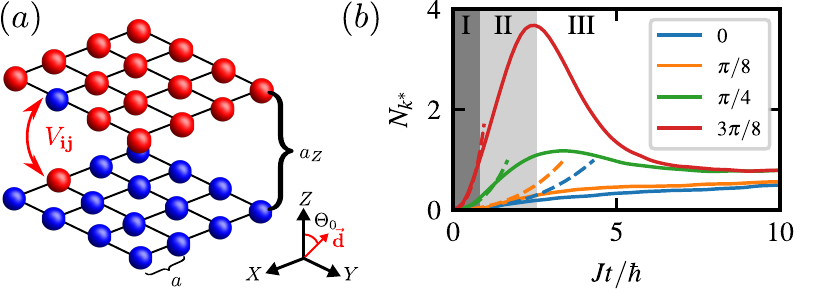} 
\caption{System. (a) Bilayer of dipoles confined in 2D planes with dipole moments aligned at an angle $\Theta_0$ to the out-of-plane direction. When the  layers are  prepared in an initial state with  opposite magnetization, dipolar inter-layer interactions create pairs of excitations in the layers in specific quasi-momentum modes. (b) Occupation of the most unstable mode $N_{k^*}$ as a function of time $t$ for different dipole orientations $\Theta_0$ (legend) at fixed $a_Z/a=2$. Shown are spin dynamics from DTWA (solid lines), and the prediction from Bogoliubov theory (dashed lines). Shaded regions indicate the regimes of dynamics (for $\Theta_0=3\pi/8$), where we find exponential growth as predicted by Bogoliubov ($\mathrm{I}$), saturation and slow-down of growth ($\mathrm{II}$), and eventual decay and thermalisation ($\mathrm{III}$). Results for a $33 \times 33$ bilayer at unit filling. \label{fig:model}}
\label{Fig1}
\end{figure} 

Here we find the exponential proliferation of correlated pairs of excitations  in spatially separated  layers. This emulates the phenomenon of pair creation from vacuum fluctuations, opening unique opportunities for quantum simulation, and for fundamental tests of quantum mechanics including  EPR steering  \cite{EPR_1935,RevModPhys.81.1727, Fadel2018,kunkel2018}.
Pair creation itself is an ubiquitous phenomenon in physics, relevant in a broad range of contexts   
including parametric amplification and two-mode squeezing in quantum optics \cite{agarwal2013quantum}, the Schwinger effect in high energy physics \cite{Schwinger_1951,Hauke_PhysRevX_2013,Kasper_PhysLettB_2016},  the emission of Unruh thermal radiation  in curved space time \cite{Unruh_1976,ChengChin_NatPhys_2019}, and in holography given that the thermofield double state generated during pair production is dual to a traversable wormhole \cite{TAKAHASHI1996,Chapman_SciPost_2019} in quantum gravity, and a resource for quantum teleportation \cite{Zhu_PNAS_2020,gao2017traversable,maldacena2017diving}.

Previous studies of pair creation processes in spinor condensates induced by contact interactions~\cite{Gross_Nature_2011,Lcke_Science_2011,Soerensen_Nature_409_2001,Qu_PRL_2020} were dominated by single (resonant) momentum modes (or trap states in confined condensates~\cite{Klempt2019}) determined by the quadratic Zeeman shift, while proposals of pair production in cavities induced by collective interactions~\cite{Sundar2022} require a set of laser tones to generate non-trivial patterns, and are sensitive to cavity loss~\cite{Periwal2021}.
In contrast, the pair creation observed in this work  allows for the generation of highly tunable,  and intriguing  distributions  of excitations   naturally emerging from anisotropic dipolar couplings \cite{Nipper2012,Yan_Nature_501_2013, Hazzard2014, dePaz_PRL_111_2013, Lepoutre2019,Patscheider_PRR_2_2020,Baranov_ChemicalReviews_112_2012,Bohn_Science_357_2017,Wall2015, moses2017new,chomaz2022dipolar}.

\sectionn{Model} %
We consider an array of frozen dipoles  with two relevant internal levels ~(e.g. two rotational states 
in the case of polar molecules) confined in two parallel two-dimensional  layers  generated via optical lattices or optical tweezers, separated by a tunable distance $a_{Z}$. We denote the upper layer as A and the lower one as B. As shown in Fig.~\ref{Fig1}(a), both layers have square geometry with a   nearest-neighbour spacing  $a$. 

Electric and magnetic dipole-dipole interactions  can lead to both  exchange of internal-state excitations, as well as Ising interactions  ~\cite{Nipper2012,Yan_Nature_501_2013, Hazzard2014, Lepoutre2019, Garbados2020, dePaz_PRL_111_2013, Patscheider_PRR_2_2020,Chen2022,Leseleuc2019}, which  
 can be tuned via  external 
electromagnetic  fields.  For the case of frozen particles,  the dynamics is governed by   the celebrated~(long-range) 
spin-1/2 XXZ model:
\begin{align}
\hat{H}_{XXZ} &= \frac{1}{2}\sum_{\sigma=A,B}\sum_{\mathbf{i}\neq \mathbf{j}}V_{\mathbf{i}\mathbf{j}}^{\sigma\sigma}
(\hat{s}_{\mathbf{i}\sigma}^{+}\hat{s}_{\mathbf{j}\sigma}^{-}+\hat{s}_{\mathbf{i}\sigma}^{-}\hat{s}_{\mathbf{j}\sigma}^{+} 
+ 2\eta \hat{s}_{\mathbf{i}\sigma}^{z}\hat{s}_{\mathbf{j}\sigma}^{z}
)  \nonumber \\
& \quad + \sum_{\mathbf{i},\mathbf{j}}V_{\mathbf{i}\mathbf{j}}^{AB}
(\hat{s}_{\mathbf{i}A}^{+}\hat{s}_{\mathbf{j}B}^{-}+\hat{s}_{\mathbf{i}A}^{-}\hat{s}_{\mathbf{j}B}^{+} 
+ 2\eta \hat{s}_{\mathbf{i}A}^{z}\hat{s}_{\mathbf{j}B}^{z}
), 
\label{Eq1}
\end{align}
where $\sigma$ indexes the layers, $\eta$ characterizes the relative strength between Ising and exchange couplings, and $\mathbf{i}=(i_{x}, i_{y})$ stands for a two-dimensional coordinate in which $i_{x}, i_{y}$ run along the positions in a given two-dimensional layer of size $N=L\times L$. As is customary, the spin operators $\hat{s}_{\mathbf{i}}^{\alpha}=\hat{\sigma}_{\mathbf{i}}^{\alpha}/2$ are given in terms of the Pauli matrices $\hat{\sigma}^{x,y,z}$ that act on the spin at site $\mathbf{i}$. We  shall focus our attention on dipole couplings of the form 
\begin{equation}
V_{\mathbf{i}\mathbf{j}}^{\sigma\sigma'} = \frac{J}{|\mathbf{r}_{\mathbf{i}}^{\sigma}-\mathbf{r}_{\mathbf{j}}^{\sigma'}
|^3}\left(1-\frac{3[\mathbf{d}\cdot(\mathbf{r}_{\mathbf{i}}^{\sigma}-\mathbf{r}_{\mathbf{j}}^{\sigma'}
)]^{2}}{|\mathbf{r}_{\mathbf{i}}^{\sigma}-\mathbf{r}_{\mathbf{j}}^{\sigma'}
|^2}\right),
\label{Eq2}
\end{equation}
where $\mathbf{\hat{d}}=\sin\Theta_{0}\hat{e}_{x}+\cos\Theta_{0}\hat{e}_{z}$ is the orientation of the dipoles, $\mathbf{r}_{\mathbf{i}}^{\sigma}$ is the position of a dipole in layer $\sigma$, and $J$ is the spin-exchange constant.

Motivated by recent experiments on 
polar molecules in bilayers~\cite{Tobias_Science_375_2022}, we consider in the following the non-equilibrium dynamics of this system  starting from an initial state where all dipoles in layer A (B) are initially in the spin up (down) state. We first analyse the spin excitations in terms of a Bogoliubov treatment, and then by simulating the quantum dynamics of the full dipolar spin model using the discrete truncated Wigner approximation (DTWA) \cite{Schachenmayer_Phys.Rev.X_5_2015,Zhu_NewJournalofPhysics_21_2019}. 

\sectionn{Bogoliubov Analysis} %
\begin{figure}[t]
\centering
\includegraphics[width=\columnwidth]{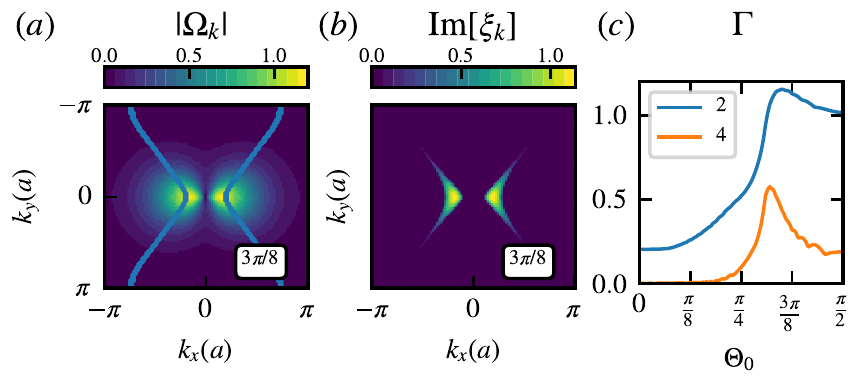}
\caption{Bogoliubov Analysis. (a) Pair coupling strength $|\Omega_k|$ as a colorplot, overlaid with the resonant surface $\varepsilon(k) \simeq 0$. (b) Imaginary part of the Bogoliubov energy $\xi_k$. Both for a dipole orientation $\Theta_0= 3\pi/8$ and $a_Z/a=2$. (c) Growth rate of the maximally unstable mode $\Gamma = \max_k \mathrm{Im}[\xi_k]$ as a function of the dipole orientation $\Theta_0$ at $a_Z/a=2,4$ as indicated in legend. All in units of $J/\hbar$. \label{fig:BDG}%
\label{}}
\end{figure}
As in the standard spin wave analysis, the spin dynamics can be described by mapping the
Hamiltonian \eqref{Eq1} to a hard-core bosonic model using the Holstein-Primakoff transformation $\hat{s}^{z}_{A,\mathbf{i}}=1/2 -\hat{a}^{\dagger}_{\mathbf{i}}\hat{a}_{\mathbf{i}}$, $\hat{s}^{+}_{A,\mathbf{i}}=\hat{a}_{\mathbf{i}}$, $\hat{s}^{-}_{A,\mathbf{i}}=\hat{a}^{\dagger}_{\mathbf{i}}$, and $\hat{s}^{z}_{B,\mathbf{i}}=-1/2+\hat{b}^{\dagger}_{\mathbf{i}}\hat{b}_{\mathbf{i}}$, $\hat{s}^{+}_{B,\mathbf{i}}=\hat{b}^{\dagger}_{\mathbf{i}}$, $\hat{s}^{-}_{B,\mathbf{i}}=\hat{b}_{\mathbf{i}}$.
The bosonic operators $\hat{a}_{\mathbf{i}}$ and $\hat{b}_{\mathbf{i}}$ characterize the spin excitations that appear on top of the prepared initial state. 

Assuming a small population of spin excitations, much smaller than the number of sites, the Hamiltonian may be rewritten in quasi-momentum space
\begin{equation}
\hat{H} = \sum_{\mathbf{k}}\tilde\varepsilon_{\mathbf{k}}(\hat{a}_{\mathbf{k}}^{\dagger}\hat{a}_{\mathbf{k}} + \hat{b}_{\mathbf{k}}^{\dagger}\hat{b}_{\mathbf{k}})+ 
\Omega_{\mathbf{k}}\hat{a}_{\mathbf{k}}^{\dagger}\hat{b}_{-\mathbf{k}}^{\dagger}+\Omega_{\mathbf{k}}^{*}\hat{b}_{-\mathbf{k}}\hat{a}_{\mathbf{k}},
\label{Eq3}
\end{equation}
 where $\hat{a}_{\mathbf{k}}=\frac{1}{\sqrt{N}}\sum_{\mathbf{r_i}}e^{-i\mathbf{k}\cdot\mathbf{r_i}}\hat{a}_{\mathbf{i}}$ and $\hat{b}_{\mathbf{k}}=\frac{1}{\sqrt{N}}\sum_{\mathbf{r_i}}e^{-i\mathbf{k}\cdot\mathbf{r_i}}\hat{b}_{\mathbf{i}}$. The momentum-dependent inter-layer coupling is given by 
$\Omega_{\mathbf{k}}=\sum_{\mathbf{j}}V_{0\mathbf{j}}^{AB}e^{-i\mathbf{k}\cdot\mathbf{r_j}}$, whereas the intra-layer band dispersion for spin excitations in each layer is $\tilde\varepsilon_{\mathbf{k}}=\varepsilon_{\mathbf{k}}-\eta(\varepsilon_{0}-\Omega_0)$, with 
$\varepsilon_{\mathbf{k}}=\sum_{\mathbf{j}\neq 0}V_{0\mathbf{j}}^{AA}e^{-i\mathbf{k}\cdot\mathbf{r_j}}$. The Ising term results in a momentum-independent shift of the intra-layer band energy.
The inter-layer coupling drives the  creation of correlated  pairs  of excitations (one  per layer) at an energy cost set by the intra-layer term.  

The Hamiltonian can be diagonalized by means of a Bogoliubov transformation \cite{supplemental}, which leads to the eigenenergies $\xi_{\mathbf{k}}  = \sqrt{\tilde\varepsilon_{\mathbf{k}}^{2}-|\Omega_{\mathbf{k}}|^{2}}$. 
Crucially, $|\Omega_{\mathbf{k}_{c}}|>|\tilde\varepsilon_{\mathbf{k}_{c}}|$ for certain quasi-momenta $\mathbf{k}_{c}$, resulting in imaginary 
eigenenergies $\xi_{\mathbf{k}_{c}}$, i.e. a dynamical instability of the vacuum of spin excitations leading to the creation of correlated pairs. The instability manifests itself as an exponential growth in the population of the corresponding mode, 
$N_{\mathbf{k}_{c}} = (|\Omega_{{k}_{c}}|/|\xi_{\mathbf{k}_{c}}|)^{2}\sinh^2{(|\xi_{\mathbf{k}_{c}}|t)}$. These predictions are shown as the dashed lines in Fig.~\ref{fig:model}(b), compared to the full spin dynamics discussed below. 

\begin{figure}
\includegraphics[width=\columnwidth]{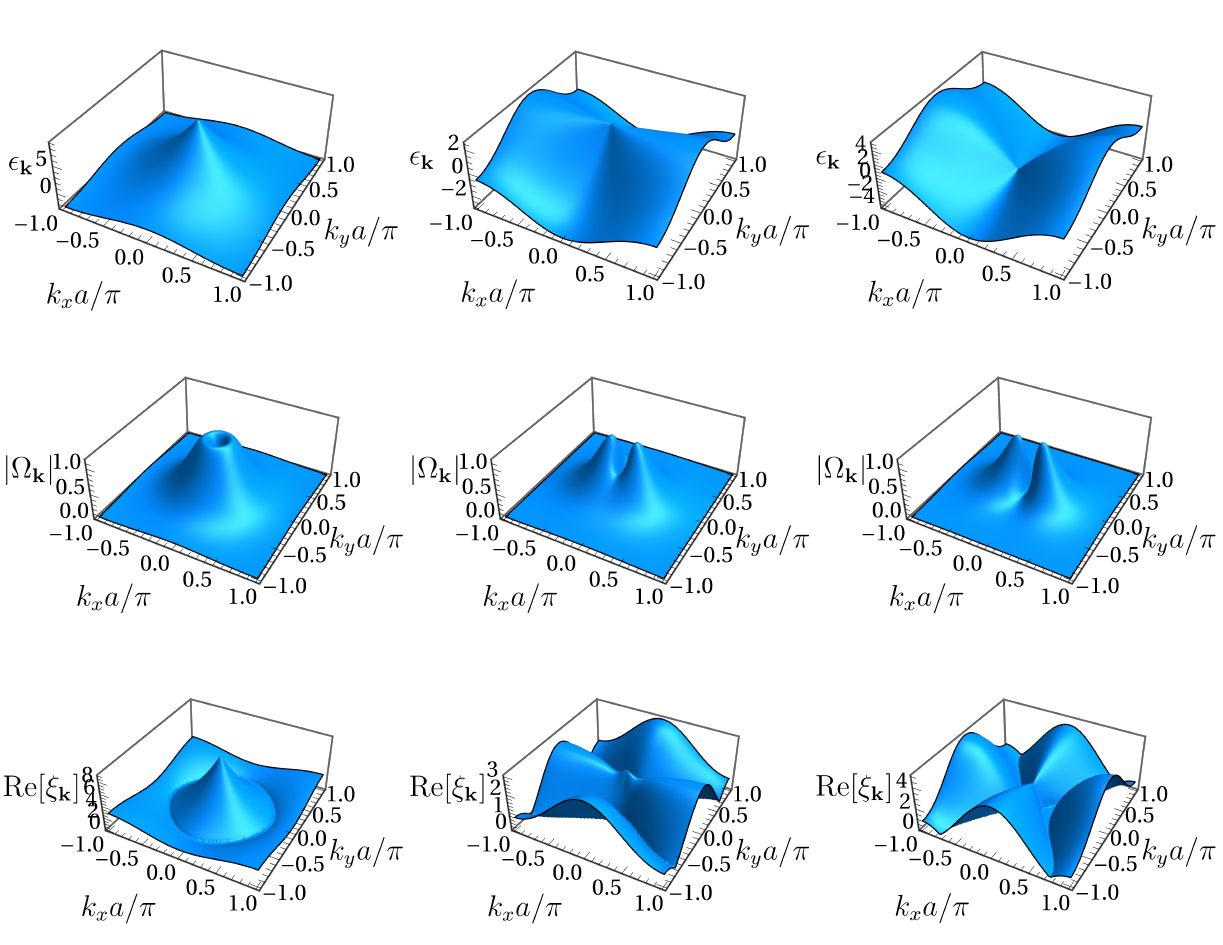}
\caption{Tunability of quasi-particle dispersions via  electric field orientation. Intra-plane dispersion $\varepsilon_{\mathbf{k}}$ (top row), inter-layer coupling $\Omega_{\mathbf{k}}$ (middle row) and real part of Bogoliubov energy $\xi_{\mathbf{k}}$ (bottom row) all in units of $J/\hbar$ for different dipole-orientations $\Theta_0 = 0, \pi/4, 3\pi/8$ (left to right columns) at a layer separation of $a_Z/a = 2$. \label{fig:BG_3D_disp}}
\end{figure}

Note that if $a_{Z}^3\gg a^3$, the inter-layer coupling 
$|\Omega_{\mathbf{k}}|$ is  much smaller than the intra-layer bandwidth. As a result, imaginary eigen-energies only occur for $\tilde\varepsilon_{\mathbf{k}} \simeq 0$. This condition is modified by the shift induced by the Ising term, which hence acts as an additional knob to tailor the quasi-momentum distributions discussed below~(a similar control knob would be provided by a layer bias of the form $\sum_i (\hat{s}^{z}_{A,\mathbf{i}}-\hat{s}^{z}_{B,\mathbf{i}})$) \cite{supplemental}. In the following we mostly focus for simplicity on the case $\eta=0$, i.e. in absence of Ising term~(XY model), for which $\tilde\varepsilon_{\mathbf{k}}=\varepsilon_{\mathbf{k}}$. 

Figure~\ref{fig:BDG}(a) shows the pair coupling strength $\Omega_{\mathbf{k}}$ in the Brillouin zone, overlaid with the resonant line for which $\varepsilon_{\mathbf{k}} \simeq 0$. Pairs are most effectively produced exactly on resonance and for momenta where pair coupling is strong. 
This is borne out in Fig.~\ref{fig:BDG}(b), which shows the growth rate of momentum modes, i.e.~the imaginary part of the Bogoliubov energy, which matches with the overlap of the resonant surface and the region of strong inter-layer coupling seen in Fig.~\ref{fig:BDG}(a). Bogoliubov theory hence predicts the creation of pairs with a specific quasi-momentum distribution. Figure~\ref{fig:BDG}(c) shows the growth rate $\Gamma$ of the most unstable mode, i.e.~the maximum of the imaginary part of the Bogoliubov energies, as a function of the dipole orientation $\Theta_0$ for two different bilayer spacings $a_Z$.  At sufficiently long times, the most unstable modes eventually dominate pair creation, resulting in vastly different dynamical scales for the spin excitations  for different dipole orientations. 
Since the overall form of the growth rate does not qualitatively change for $a_Z^3 \gg a^3$, we will focus on the case $a_Z/a=2$.

\sectionn{Tunability and control over unstable modes} %
%
Next we illustrate the tunability of and control over the momentum structure of the dynamical instability, which controls the spatial structure of the created pair correlations and entanglement, and the growth rate of the most unstable modes, as well as the topology of the unstable modes. In particular, we demonstrate how the topology changes from a simply-connected circular manifold at $\Theta_0=0$, to two disconnected arcs above a critical $\Theta_0$.

There are different natural parameters that allow us to tune the pair creation instabilities. The orientation of the dipoles $\Theta_0$ via the field direction, or a shift of the intra-layer dispersion $\tilde\varepsilon_{\mathbf{k}}$, which may be induced either by an Ising term $\hat{s}_i^z \hat{s}_j^z$ in the dipolar interactions present at finite electric fields or by a layer bias of the form $h \sum_i (\hat{s}^{z}_{A,\mathbf{i}}-\hat{s}^{z}_{B,\mathbf{i}})$ induced by an electric field gradient.

\begin{figure}
\includegraphics[height=0.5\columnwidth]{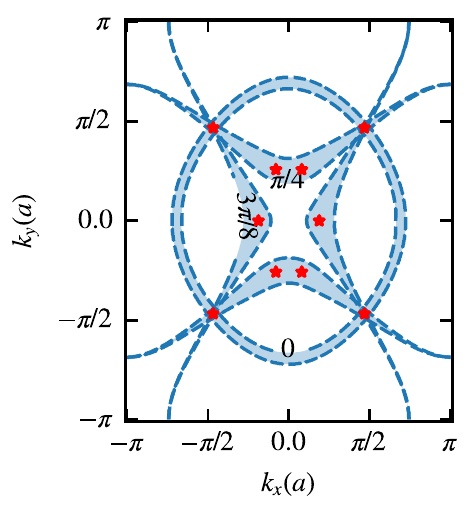}
\includegraphics[height=0.5\columnwidth]{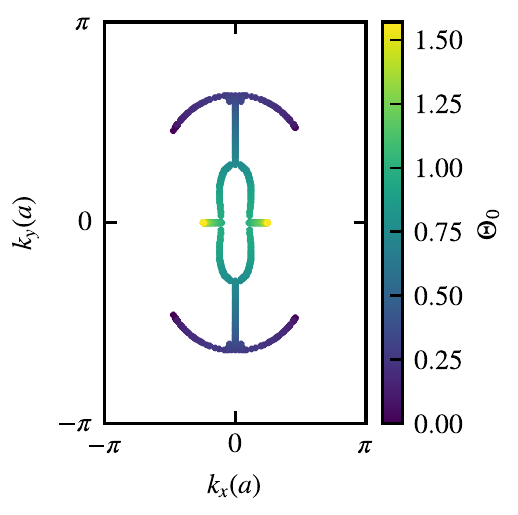}
\caption{Tunability of unstable modes via electric field orientation. Left panel. Manifolds of unstable modes defined as non-vanishing imaginary part of the Bogoliubov energy $\xi_{\mathbf{k}}$. Red stars indicate the most unstable mode.  Right panel. Contours of most unstable mode $\mathbf{k}_*$  as a function of dipole orientation $\Theta_0$ (color bar). \label{fig:orientation_tunability_2}}
\end{figure}

We begin by illustrating the effect of changing the dipole orientation in Fig.~\ref{fig:BG_3D_disp}. We note the intra-layer dispersion (top row) shows a Dirac cone-like structure at $\Theta_0=0$, with linear scaling around $k=0$ with an almost rotationally  symmetric dispersion. Whereas at any finite $\Theta_0$ the dispersion  becomes strongly anisotropic. Similar behavior is seen in the inter-layer coupling $\Omega_{\mathbf{k}}$ in the middle row. We also note that for the chosen layer-spacing $a_Z/a = 2$, the intra-layer dispersion is significantly larger than the pair-coupling. Since the Bogoliubov dispersions (bottom row) is given by $\xi_{\mathbf{k}}  = \sqrt{\tilde\varepsilon_{\mathbf{k}}^{2}-|\Omega_{\mathbf{k}}|^{2}}$ we find unstable modes with zero real part close to regions of vanishing dispersion only. We also emphasise that the unstable modes form a ring-like structure for $\Theta_0=0$, whereas the unstable modes form two separate arc-like features for the other orientations. 

We separately illustrate the full tunability of the manifolds of unstable modes through distinct topologies via the dipole orientation in Fig.~\ref{fig:orientation_tunability_2}. We observe both a change from a connected circular structure to separate arcs, as well as a change in the number of most unstable modes, from 2 to 4, for different dipole orientations.

\begin{figure}
\includegraphics[width=\columnwidth]{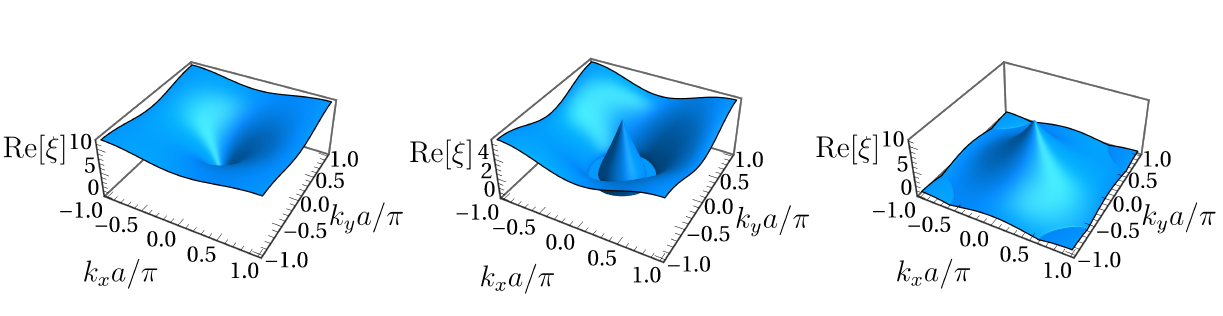}
\includegraphics[height=0.5\columnwidth]{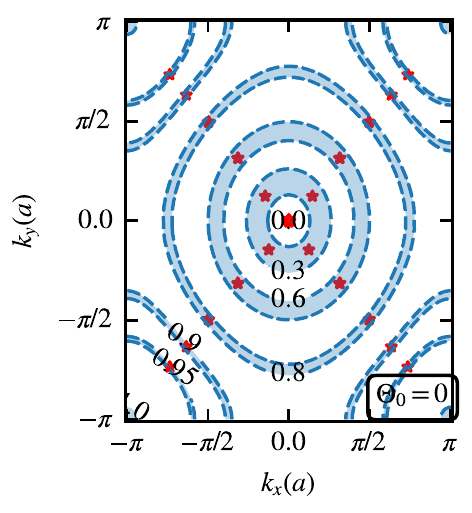}
\includegraphics[height=0.5\columnwidth]{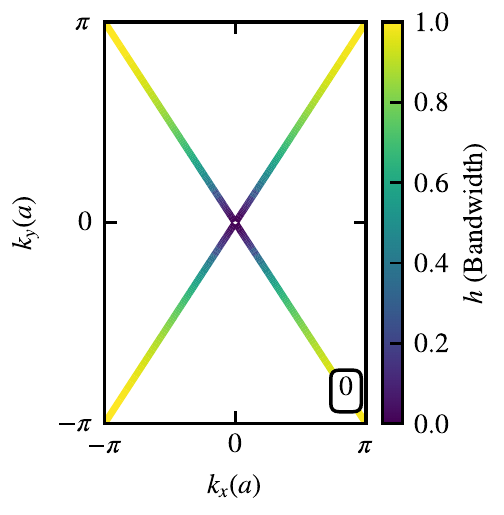}
\caption{Tunability of unstable modes via electric field gradient at $\Theta_0=0$. Top panel. Real part of the Bogoliubov quasi-energy $\xi_{\mathbf{k}}$ in units of $J/\hbar$ at $\Theta_0=0.0$  and $a_Z/a =2$ for different applied field gradients $x=0,0.5,0.95$ from left to right. The applied field gradient is defined via a fraction $x$ of the total bandwidth $W= \max \varepsilon_{\mathbf{k}} - \min \varepsilon_{\mathbf{k}}$ as $h = x \, W +h_0$ where $h_0$ is chosen to shift the dispersion to have $\tilde \varepsilon_{\mathbf{0}}=0$.  Bottom left panel. Manifolds of unstable modes defined as non-vanishing imaginary part of the Bogoliubov energy $\xi_{\mathbf{k}}$. Red stars indicate the most unstable mode.  Bottom right panel. Contours of most unstable mode $\mathbf{k}_*$  as a function of the applied field gradient (color bar).\label{fig:bias_tunability_1}}
\end{figure}


In addition, we may also tune the instability by a shift of the dispersion. For simplicity, we consider an applied electric  field gradient  as it is in principle fully tunable. We define the layer bias $h = x \, W +h_0$ as a fraction $x$ of the bandwidth of the intra-layer dispersion  $W= \max \varepsilon_{\mathbf{k}} - \min \varepsilon_{\mathbf{k}}$ with an additional offset $h_0$ to shift the dispersion to have $\tilde \varepsilon_{\mathbf{0}}=0$. We show in Fig.~\ref{fig:bias_tunability_1} how this allows control over the unstable manifolds from a single point at $k_c=(0,0)$ for $x=0$ (left top panel) over ring-like structures around the centre of the Brillouin zone at intermediate $x$ (central top panel), to arcs around the corners of the Brillouin zone (BZ)  as $x$ approaches  1 (top right panel).  The instabilities  are pushed to the  4 corners of the BZ at $x=1$. This is shown more directly in the right bottom panel of  Fig.~\ref{fig:bias_tunability_1} which shows the behavior of the most unstable mode as a function of the layer bias. 

We note that for $\Theta_0=0$ the most unstable mode is always four-fold degenerate for $\Theta_0=0$, with the exception of the case of $k \rightarrow 0$ and close to the  degenerate region  along rings around the centre of the BZ. In contrast, for other dipole orientations, e.g. $\Theta_0=\pi/4$ (not shown), the most unstable mode may be either two or four-fold degenerate and of the form $\mathbf{k}_* = \pm \mathbf{k}$ or $\mathbf{k}_* = (\pm k_x,\pm k_y)$. In addition we observe another distinct topology of unstable modes consisting of two separate disk-like regions.

\sectionn{Full spin dynamics} %
\begin{figure}
\includegraphics[width=\columnwidth]{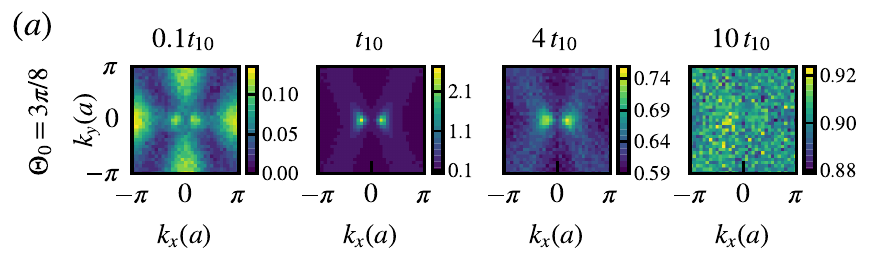}
\includegraphics[width=\columnwidth]{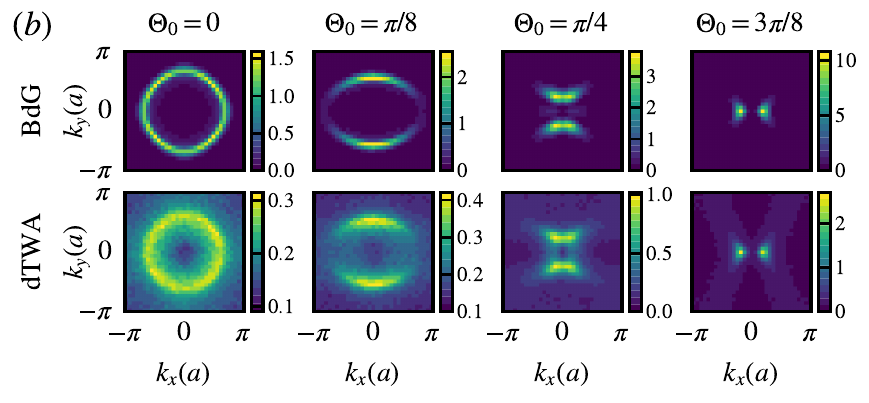}
\caption{Momentum structure of created pairs $N_\mathbf{k}$. %
(a) Time evolution of $N_\mathbf{k}(t)$ within DTWA showing the different regimes of dynamics for $\Theta_0=3 \pi/8$. Time in terms of $t_{10}$, where $N_{\textrm{pair}}(t_{10}) = 0.1 N$. The left most panel shows the early time regime before  exponential growth has taken over. The second panel shows the build up of the expected momentum structure. The last two panels show the subsequent thermalisation as scattering between momentum modes occurs. %
(b) Comparison of Bogoliubov prediction (top) and spin dynamics from DTWA (bottom) for a range of $\Theta_0$, all at $t_{10}$.
Results for a $33 \times 33$ bilayer with layer spacing $a_Z/a=2$ with open boundary conditions  at unit filling.\label{fig:momentum_structure}}
\end{figure} 
We next turn to the full quantum spin dynamics of the model obtained within the DTWA~\cite{Schachenmayer_Phys.Rev.X_5_2015,Zhu_NewJournalofPhysics_21_2019}. The momentum state population of excitations in layer B maps  to the  structure factor which in terms of spin-operators can be written  as
\begin{equation}
\hat{N}^B_{\mathbf{k}} = \frac{1}{N}\sum_{\mathbf{i}\mathbf{j}} e^{ i \mathbf{k} \cdot (\mathbf{r}_{\mathbf{i}} -\mathbf{r}_{\mathbf{j}} )} \hat{b}_{\mathbf{i}}^{\dagger} \hat{b}_{\mathbf{j}} = \frac{1}{N}\sum_{\mathbf{i}\mathbf{j}} e^{ i \mathbf{k} \cdot(\mathbf{r}_{\mathbf{i}} -\mathbf{r}_{\mathbf{j}} )} \hat{s}_{\mathbf{i}}^{+} \hat{s}_{\mathbf{j}}^-
\end{equation}
with a similar expression for layer $A$ and we define $N_{\mathbf{k}}= \langle \hat{N}^A_{\mathbf{k}} \rangle= \langle \hat{N}^B_{\mathbf{k}} \rangle$.  Note that the disconnected part vanishes identically \cite{supplemental}. We will focus on this momentum structure to observe the pair creation process in the spin dynamics (see \cite{supplemental} for real-space results).


Figure~\ref{fig:model}(b) shows the population $N_{k^*}(t)$ of the most unstable mode $k^*$ for different dipole orientations $\Theta_0$ obtained from both DTWA simulations (solid lines) and the Bogoliubov analysis (dashed lines) with no fitting parameters. Both results are in very good agreement in the initial exponential growth regime (regime $\mathrm{I}$), in which a  significant number of pairs are created before corrections or further scattering terms become important. This is followed in the full dynamics by a slow down and eventual saturation to a maximal mode occupation (regime $\mathrm{II}$), after which scattering between momentum modes starts to deplete the maximally unstable mode ($\mathrm{III}$). As expected from the Bogoliubov analysis, we observe that the spatial  and temporal growth of correlations  exhibit  a strong dependence  on the dipole orientation.

We show the time evolution of the full momentum distribution of the created pairs during the spin dynamics of the model, obtained within DTWA for a representative $\Theta_0=3 \pi/8$ in Fig.~\ref{fig:momentum_structure}(a), with an extended set of figures provided in the SI \cite{supplemental}. At very short times off-resonant non exponentially growing modes dominate the structure (left panel), which then give way to the exponentially growing unstable modes resulting in the distribution expected from the Bogoliubov prediction (second panel). Naturally, higher-order terms neglected within the Bogoliubov approximation will eventually result in scattering between different momentum modes leading to thermalisation. This expectation is seen in the last two panels showing first an increase of population in the slower growing unstable modes and then thermalisation in the late time regime. We note that the approach to equilibrium can itself host rich physics \cite{Pruefer_Nature_2018,Erne_Nature_2018,Glidden_NatPhys_2021,RodriguezNieva2022}

We establish the correspondence of the DTWA results and the Bogoliubov predictions for different dipole orientations $\Theta_0$ in Fig.~\ref{fig:momentum_structure}(b). Here, we choose an evolution time $t$ such that the total number of pairs $N_{\mathrm{pair}}=\sum_{\mathbf{k}} N_{\mathbf{k}}(t) = 0.1 N$, to allow time for the dynamical instability to create pairs, while at the same time keeping within the regime of validity of the Bogoliubov analysis. 
We observe good agreement for all dipole orientations indicating that the pair production mechanism is still effective in the full dipolar spin model. 

\sectionn{Wider context}
After establishing these phenomena in the full spin dynamics, we can now connect back to the motivating ideas. Conceptually, we realise quantum-time evolution of the form $|\psi(t)\rangle \simeq e^{\Gamma t \sum_{{\bf k} \in\eta_c} \beta_{{\bf k}}^\dag \beta_{{\bf k}}} |{\mathrm {vac}}\rangle$, where $|{\mathrm {vac}}\rangle$ is the vacuum of excitations, $\eta_c$ is the set of most unstable momentum modes and $\beta_{{\bf k}}$ are the quasiparticle operators. This generates pairs of correlated excitations with opposite momenta in the layers, the momentum distribution at momentum $\bf k$ in layer A equals that at momentum $-\bf k$ in layer B which reflects the strongly entangled character of the state generated during pair creation. 

As a direct consequence of realising a pair-creation Hamiltonian of momentum modes, the quantum dynamics maps onto Unruh radiation \cite{ChengChin_NatPhys_2019}: the created population of excitations in momentum space is exactly the thermal bath observed in an accelerated frame \cite{supplemental}. Moreover, the time evolved state \cite{supplemental} is the thermofield double state \cite{TAKAHASHI1996,Chapman_SciPost_2019}, which within the  the holographic correspondence, is  dual to wormholes on the gravity side \cite{israel1976thermo,maldacena2003eternal} and enables teleportation (“traversable wormholes”) \cite{Zhu_PNAS_2020,gao2017traversable,maldacena2017diving}. It is also an ideal resource for  EPR steering \cite{EPR_1935,RevModPhys.81.1727, Fadel2018,kunkel2018} given its shared entanglement between spatially separated layers. Finally, the entangled pairs in the form of a two-mode squeezed state \cite{agarwal2013quantum} also feature correlations in the amplitude and phase quadratures of the momentum modes: their individual fluctuations are amplified, while their relative fluctuations are reduced below the vacuum noise level enabling applications in quantum metrology.

\sectionn{Imperfect filling} %
\begin{figure}
\includegraphics[width=\columnwidth]{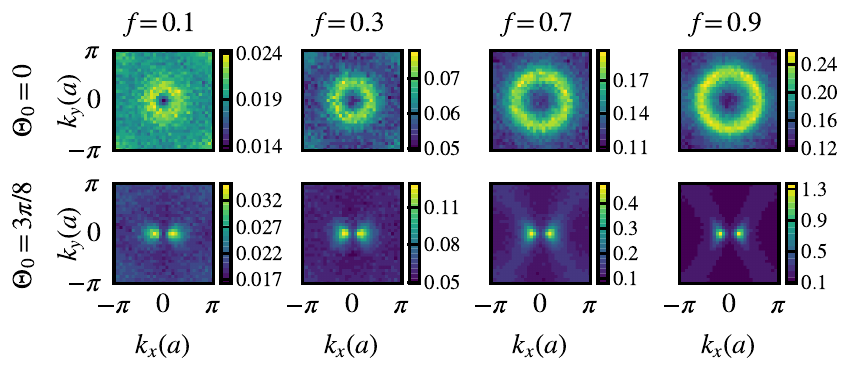}
\caption{Momentum occupation $N_{\mathbf{k}}$ of created pairs in presence of positional disorder/non-unit filling obtained within DTWA. Results for a range of $\Theta_0$ and filling fraction $f$ at fixed $a_Z/a=2$ at times such that $N_{\textrm{pair}}(t)= 0.1 f N$. $L=33$ with open boundary conditions.\label{fig:momentum_structure_disorder}}
\end{figure} 
Considering the feasibility to observe these effects in an experimental setting, while tweezer arrays offer the possibility to achieve unit filling \cite{Endres_Science_2016,Browaeys_Science_2016,NatPhysKaufman2021,Zhang2022,holland2022ondemand,bao2022dipolar,Anderegg2019}, a major challenge in optical lattices, especially for polar molecules,  is imposed  by imperfect filling, which results in positional disorder of the pinned dipoles.
While  important  developments in cooling and trapping molecules  have allowed the preparation of lattice arrays with up  to $f=0.25$~\cite{demarco2019degenerate,Valtolina_2020,moses2015creation,christakis2022probing},  which    highlight  the near-future potential of achieving high filling fractions, they  also illustrate the need to understand which effects would be observable at lower filling fractions in current setups.

%

To address this question, we consider bilayers in which each lattice site has a fixed spatially uniform probability $f$ to be occupied or empty. We show the resulting momentum occupation $N_k$ for different filling fractions $f$~(averaging over $10000$ filling realizations) and two dipole orientations $\Theta_0$ in Fig.~\ref{fig:momentum_structure_disorder}. Also for the case of imperfect filling our DTWA results are in very good agreement with the Bogoliubov analysis~(for more details see~\cite{supplemental}). We observe that while the signal to noise deteriorates as the lattice becomes more sparsely filled, most importantly, the main qualitative phenomenology, the emergence of a manifold of unstable exponentially growing modes, does extend to  a remarkably low filling fraction regime, which makes the observation in experimental platforms feasible.

\sectionn{Outlook} %
Dipolar systems confined in two-dimensional bilayers host a dynamical instability generating correlated pairs and entanglement between spatially separated layers. Making use of the wide tunability of dipolar interactions, one can access different shapes and topologies of the momentum distribution of the created pairs. These correlations may be probed using spatially-resolved measurements accessible in state-of-the-art platforms in tweezers and quantum gas microscopes for a range of atomic or molecular gases. In   these systems, the entangled pairs in spatially separated layers can be stored and manipulated with individual particle control  providing unique new opportunities in metrology.

The reported  dynamical instabilites  not only are genuinely driven by quantum fluctuations but in stark contrast to usual roton-like instabilities at finite momentum, as e.g. in dipolar condensates~\cite{Chomaz2018}, here the generated modes of opposite momentum $\pm \vec k$ are not equivalent, but rather correspond to two different entangled pairs of excitations in both layers.

The long-time behaviour and eventual thermalisation of the excitations remain an open question. Since the initial pairing instability creates a well-defined highly non-thermal occupation in momentum space, the eventual approach to equilibrium might reveal universal non-equilibrium scaling exponents and self-similiarity \cite{Pruefer_Nature_2018,Erne_Nature_2018,Glidden_NatPhys_2021,RodriguezNieva2022}.

\begin{acknowledgments}
\noindent{\textit{Acknowledgements:}
We  acknowledge careful review of this manuscript and  useful  comments  from A. Carroll and J. Higgins. G.A.D.-C. and L.S. acknowledge support of the Deutsche Forschungsgemeinschaft (DFG, German Research Foundation) under Germany's Excellence Strategy -- EXC-2123 QuantumFrontiers -- 390837967.
D.W and A.M.R. acknowledge support from  the AFOSR MURI,  the ARO single investigator award W911NF-19-1-0210,   the  NSF JILA-PFC PHY-1734006 grants, NSF QLCI-2016244 grants, by the DOE Quantum Systems Accelerator (QSA) grant and by NIST.}
\end{acknowledgments}

\nocite{supplemental}
\bibliography{pair_creation}{}

\appendix

\maketitle
\setcounter{equation}{0}
\setcounter{figure}{0}
\setcounter{table}{0}
\makeatletter
\renewcommand{\theequation}{S\arabic{equation}}
\renewcommand{\thefigure}{S\arabic{figure}}

%
\setcounter{equation}{0}
\setcounter{figure}{0}
\setcounter{table}{0}
\makeatletter
\renewcommand{\thefigure}{S\arabic{figure}}
\section{Appendix}
The appendices contains additional details on the Bogoliubov analysis, the effect of boundary conditions,  the real space structure of the spin correlations, and Bogoliubov as well as extended DTWA results for the finite filling fraction behavior of the momentum structure of correlations, and extended DTWA results for the time-dependence.

\subsection{Bogoliubov Analysis}
In this section, we provide further details on the
Bogoliubov analysis of the spin Hamiltonian. First, we focus on the diagonalization procedure of a unit filling lattice, then we proceed to discuss the case of lattices with fillings smaller than one. For a perfectly filled lattice, we may write the Hamiltonian in quasi-momentum space:\begin{equation}
\begin{split}
\hat{H} &= \sum_{\mathbf{k}}\varepsilon_{\mathbf{k}}(\hat{a}_{\mathbf{k}}^{\dagger}\hat{a}_{\mathbf{k}} + \hat{b}_{\mathbf{k}}^{\dagger}\hat{b}_{\mathbf{k}})+\\
& \sum_{\mathbf{k}}| \left [\Omega_{\mathbf{k}}|e^{-i\alpha_{\mathbf{k}}}\hat{a}_{\mathbf{k}}^{\dagger}\hat{b}_{-\mathbf{k}}^{\dagger}+|\Omega_{\mathbf{k}}|e^{i\alpha_{\mathbf{k}}}\hat{a}_{\mathbf{k}}\hat{b}_{-\mathbf{k}} \right ],
\end{split}
\label{}
\end{equation}
where we have made explicit the complex nature of the inter-layer coupling $\Omega_{\mathbf{k}}=|\Omega_{\mathbf{k}}|e^{-i\alpha_{\mathbf{k}}}$. In the presence of an Ising term, we would simply substitute $\varepsilon_{\mathbf{k}}$ by $\tilde \varepsilon_{\mathbf{k}}$. Before introducing the Bogoliubov transformation it is convenient to decouple the above Hamiltonian into symmetric and antisymmetric collective quasi-momentum modes. For this purpose, we define the following operators 
\begin{equation}
\begin{split}
\hat{S}_{\mathbf{k}} &= \frac{1}{\sqrt{2}}(e^{-i\alpha_{\mathbf{k}}/2}\hat{a}_{\mathbf{k}}+e^{i\alpha_{\mathbf{k}}/2}\hat{b}_{\mathbf{k}})\\
\hat{A}_{\mathbf{k}} &= \frac{1}{\sqrt{2}}(e^{-i\alpha_{\mathbf{k}/2}}\hat{a}_{\mathbf{k}}-e^{i\alpha_{\mathbf{k}}}\hat{b}_{\mathbf{k}}).
\end{split}
\end{equation}
In terms of these new operators, the Hamiltonian can be rewritten as $\hat{H}=\hat{H}_{S}+\hat{H}_{A}$ with
\begin{equation}
\begin{split}
\hat{H}_{S} &= \sum_{\mathbf{k}}\varepsilon_{\mathbf{k}}\hat{S}_{\mathbf{k}}^{\dagger}\hat{S}_{\mathbf{k}}+\frac{|\Omega_{\mathbf{k}}|}{2}(\hat{S}_{\mathbf{k}}^{\dagger}\hat{S}_{-\mathbf{k}}^{\dagger}+\hat{S}_{\mathbf{k}}\hat{S}_{-\mathbf{k}})\\
\hat{H}_{A} &= \sum_{\mathbf{k}}\varepsilon_{\mathbf{k}} 
\hat{A}_{\mathbf{k}}^{\dagger}\hat{A}_{\mathbf{k}}-\frac{|\Omega_{\mathbf{k}}|}{2}(\hat{A}_{\mathbf{k}}^{\dagger}\hat{A}_{-\mathbf{k}}^{\dagger}+\hat{A}_{\mathbf{k}}\hat{A}_{-\mathbf{k}}).
\end{split}
\end{equation}
In the following, we discuss the diagonalization of $\hat{H}_{S}$, 
but that of $\hat{H}_{A}$ is completely analogous. At this point, we introduce the Bogoliubov transformation $\hat{\beta}_{\mathbf{k}}=u_{\mathbf{k}}\hat{S}_{\mathbf{k}}-v_{\mathbf{k}}^{*}\hat{S}_{-\mathbf{k}}^{\dagger}$. 
The amplitudes $u_{\mathbf{k}}$ and $v_{\mathbf{k}}$ obey the 
Bogoliubov-de Gennes equations:
\begin{eqnarray}
\xi_{\mathbf{k}} u_{\mathbf{k}} &=&  \varepsilon_{\mathbf{k}} u_{\mathbf{k}} + |\Omega_{\mathbf{k}}| v_{\mathbf{k}}, \\
\xi_{\mathbf{k}} v_{\mathbf{k}} &=&  -|\Omega_{\mathbf{k}}| u_{\mathbf{k}}-\varepsilon_{\mathbf{k}} v_{\mathbf{k}},
\end{eqnarray}
where the eigenenergies acquire the form
$\xi_{\mathbf{k}}=\sqrt{\varepsilon_{\mathbf{k}}^{2}-|\Omega_{\mathbf{k}}|^{2}}$. In the case of real eigenvalues, the time dependence of the Bogoliubov operators is $\hat{\beta}_{\mathbf{k}}(t) = e^{-i\xi_{\mathbf{k}}t}\hat{\beta}_{\mathbf{k}}(0)$ and $\hat{\beta}_{\mathbf{k}}^{\dagger}(t) = e^{i\xi_{\mathbf{k}}t}\hat{\beta}_{\mathbf{k}}^{\dagger}(0)$. Inversion of the Bogoliubov transformation yields the following expression
\begin{equation}
\begin{split}
\hat{S}_{\mathbf{k}}(t) &= [e^{-i\xi_{\mathbf{k}}t}\cosh^{2} \phi_{\mathbf{k}} -e^{i\xi_{\mathbf{k}}t}\sinh^{2}{\phi_{\mathbf{k}}}]\hat{S}_{\mathbf{k}}(0) + \\
&i\sinh{(2\phi_{\mathbf{k}})}\sin(\xi_{\mathbf{k}}t)\hat{S}_{\mathbf{k}}^{\dagger}(0),
\end{split}
\label{}
\end{equation}
with $\sinh^{2}{2\phi_{\mathbf{k}}} = |\Omega_{\mathbf{k}}|^{2}/\xi_{\mathbf{k}}^{2}$. The vacuum expectation value of the population of the symmetric mode $\mathbf{k}$ gives $\langle 0|\hat{S}_{\mathbf{k}}^{\dagger}(t)\hat{S}_{\mathbf{k}}(t)|0\rangle = \sinh^{2}{(2\phi_{\mathbf{k}})\sin^{2}{(\xi_{\mathbf{k}}t)}}$, the same expression fulfills $\langle 0|\hat{A}_{\mathbf{k}}^{\dagger}(t)\hat{A}_{\mathbf{k}}(t)|0\rangle$. Then, the total population of the mode is simple
\begin{equation}
\begin{split}
N_{\mathbf{k}} &= \langle\hat{A}_{\mathbf{k}}^{\dagger}(t)\hat{A}_{\mathbf{k}}(t)+\hat{S}_{\mathbf{k}}^{\dagger}(t)\hat{S}_{\mathbf{k}}(t)\rangle/2\\ &= [|\Omega_{\mathbf{k}}|\sin{(\xi_{\mathbf{k}}t)}/\xi_{\mathbf{k}}]^{2}
\label{AppEq5}
\end{split}
\end{equation}
If $\xi_{\mathbf{k}}$ is imaginary, the Bogoliubov modes fulfill $|u_{\mathbf{k}}|^{2}=|v_{\mathbf{k}}|^{2}$ and therefore the modes are actually quadratures of the form
\begin{equation}
\begin{split}
\hat{X}_{\mathbf{k}} &=\frac{1}{\sqrt{\sin\phi_{\mathbf{k}}}}\left[e^{-i\phi_{\mathbf{k}}/2}\hat{S}_{\mathbf{k}}-e^{i\phi_{\mathbf{k}}/2}\hat{S}_{-\mathbf{k}}^{\dagger}\right]\\ 
\hat{P}_{\mathbf{k}} &= \frac{1}{\sqrt{\sin\phi_{\mathbf{k}}}}\left[e^{i\phi_{\mathbf{k}}/2}\hat{S}_{\mathbf{k}}-e^{-i\phi_{\mathbf{k}}/2}\hat{S}_{-\mathbf{k}}^{\dagger}\right],
\end{split}    
\end{equation}
with $\tan \phi_{\mathbf{k}} = -\varepsilon_{\mathbf{k}}/|\xi_{\mathbf{k}}|$. The first quadrature grows exponentially in time $\hat{X}_{\mathbf{k}}(t) = e^{|\xi_{\mathbf{k}}|t}\hat{X}_{\mathbf{k}}(0)$, whereas $\hat{P}_{\mathbf{k}}(t) = e^{-|\xi_{\mathbf{k}}|t}\hat{P}_{\mathbf{k}}(0)$ decreases exponentially. By inverting the definition of the quadratures one can find the time evolution of the symmetric mode
\begin{equation}
\begin{split}
\hat{S}_{\mathbf{k}} &= \frac{i}{\sqrt{2\sin\phi_{\mathbf{k}}}}[(e^{-i\phi_{\mathbf{k}}}e^{|\xi_{\mathbf{k}}|t} -e^{i\phi_{\mathbf{k}}}e^{-|\xi_{\mathbf{k}}|t})\hat{S}_{\mathbf{k}}(0)\\
& -2\sinh{(|\xi_{\mathbf{k}}|t)}\hat{S}_{-\mathbf{k}}^{\dagger}(0)]
\end{split}
\label{}
\end{equation}
then it follows that $\langle\hat{S}_{\mathbf{k}}^{\dagger}(t)\hat{S}_{\mathbf{k}}(t)\rangle = \sinh^{2}{(|\xi_{\mathbf{k}}|t)}/\sin^{2}\phi_{\mathbf{k}}^{2}$, a similar expression fulfills $\langle\hat{A}_{\mathbf{k}}^{\dagger}(t)\hat{A}_{\mathbf{k}}(t)\rangle$. The total population of the mode is simple $N_{\mathbf{k}}  = [|\Omega_{\mathbf{k}}|\sinh{(|\xi_{\mathbf{k}}|t)}/|\xi_{\mathbf{k}}|]^{2}$.
Since $\sin{(i|\xi_{\mathbf{k}}|)}/i|\xi_{\mathbf{k}}|\rightarrow -\sinh{(|\xi_{\mathbf{k}}|)}/|\xi_{\mathbf{k}}|$, one can safely use the expression in Eq. (\ref{AppEq5}) to obtain the time dependence of the density of excitations in each layer
\begin{equation}
    n(t)a^{2} = \int_{BZ} \frac{d^{2}k}{(2\pi)^{2}}|\Omega_{k}|^{2}\left[\frac{\sin{|\xi_{\mathbf{k}}|t}}{|\xi_{\mathbf{k}}|} \right],
\label{}
\end{equation}
where the integration is over the first Brillouin zone.

The Bogoliubov treatment of the case of imperfect filling is more involved. We consider a lattice with $L\times L$ sites with open boundary conditions, and a filling $f<1$. 
We create a given realization by randomly filling each layer 
with a given number of dipoles, up to the desired lattice filling. Due to positional disorder, it is suitable to work with the Hamiltonian in space representation
\begin{equation}
\begin{split}
\hat{H} &= \sum_{\mathbf{i}\neq\mathbf{j}}V_{\mathbf{i}\mathbf{j}}^{AA}\hat{a}_{\mathbf{i}}^{\dagger}\hat{a}_{\mathbf{j}}+\sum_{\mathbf{i}\neq\mathbf{j}}V_{\mathbf{i}\mathbf{j}}^{BB}\hat{b}_{\mathbf{i}}^{\dagger}\hat{b}_{\mathbf{j}}\\
& + \sum_{\mathbf{i},\mathbf{j}}V_{\mathbf{i}\mathbf{j}}^{AB}\hat{a}_{\mathbf{i}}^{\dagger}\hat{b}_{\mathbf{j}}^{\dagger}+\sum_{\mathbf{i},\mathbf{j}}V_{\mathbf{i}\mathbf{j}}^{BA}\hat{b}_{\mathbf{i}}\hat{a}_{\mathbf{j}}.
\end{split}
\end{equation}
We may again apply the Bogoliubov transformation, $\hat{\beta}_{n}=\sum_{\mathbf{j}}u_{n\mathbf{j}}\hat{a}_{\mathbf{j}}+\sum_{\mathbf{j}'}v_{n\mathbf{j}'}\hat{b}_{\mathbf{j}'}$.
By imposing that $\xi_{n}\hat{\beta_{n}}=[\hat{\beta}_{n}, \hat{H}]$, we obtain the Bogoliubov-de-Gennes equations
\begin{equation}
\xi_{n}
\begin{pmatrix}
\mathbf{u}_{n}\\
\mathbf{v}_{n}
\end{pmatrix}
=
\begin{pmatrix}
V^{AA} & -V^{AB}\\
V^{BA} & -V^{BB}
\end{pmatrix}
\begin{pmatrix}
\mathbf{u}_{n}\\
\mathbf{v}_{n}
\end{pmatrix}
,
\end{equation}
where $\mathbf{u}_{n} = (u_{n,\mathbf{i}_{1}}, u_{n,\mathbf{i}_{2}} \dots  u_{n,\mathbf{i}_{L\times L}})^{T}$ and similarly for $\mathbf{v}_{n}$. By solving the above eigenvalue problem, we obtain the eigenmodes and their corresponding evolution in time. Inverting the Bogoliubov transformation provides the time dependence of the lattice operators, and Fourier transforming yields the quasi-momentum distribution. Averaging over many random realizations of the lattice filling, we obtain the distributions discussed below.

\subsection{Pair Creation for quantum simulation and metrology}
In this section we provide additional details on the relevance of the discussed pair-creation mechanism to quantum simulation and metrology.

At the centre of these connections is that we effectively realise pair-creation at resonant momenta, or a two-mode squeezing Hamiltonian involving  momentum modes. This  allows for the generation of the  so called  thermofield double states (TFD). TFDs are not only at the heart of  quantum simulation of Unruh radiation \cite{ChengChin_NatPhys_2019}, but also  a resource for   entanglement generation between spatially separated modes  which can be used for various applications ranging from  quantum metrology, over teleportation to quantum communication.

To make these connection as transparent as possible we restrict for simplicity to a single resonant mode, for which we effectively realise
\begin{equation}
  h_{\mathbf{k}_c} =   \left(  \Omega_{\mathbf{k}_c} \hat{a}^{\dagger}_{\mathbf{k}_c} \hat{b}^{\dagger}_{-\mathbf{k}_c}+\Omega_{\mathbf{k}_c}^* \hat{a}_{-\mathbf{k}_c} \hat{b}_{\mathbf{k}_c} \right)
\end{equation}
which is the well known two-mode squeezing Hamiltonian, which here creates entangled pairs in spatially separated layers A and B. \\

The operators consequently evolve as
\begin{align}
\begin{pmatrix} \hat{a}_{k} (t) \\
                             \hat{b}^{\dagger}_{-k} (t)\\
\end{pmatrix}
 = 
 \begin{pmatrix} \cosh(|\Omega_{\mathbf{k}_c}| t/\hbar) & \sinh(|\Omega_{\mathbf{k}_c}| t/\hbar)  \\
                 \sinh(|\Omega_{\mathbf{k}_c}| t/\hbar)  & \cosh(|\Omega_{\mathbf{k}_c}| t/\hbar)  \\
\end{pmatrix}
\begin{pmatrix} \hat{a}_{k} (0) \\
                             \hat{b}^{\dagger}_{-k} (0)\\
\end{pmatrix}
\end{align}

One can compare this directly to the transformation into the Rindler frame of a scalar field \cite{ChengChin_NatPhys_2019}, which is given by
\begin{align}
\begin{pmatrix} \hat{b}_{\omega}^R  \\
                \hat{b}^{\dagger L}_{\omega}\\
\end{pmatrix}
 = 
 \begin{pmatrix} \cosh(r_\omega) & \sinh(r_\omega)  \\
                 \sinh(r_\omega)  & \cosh(r_\omega)  \\
\end{pmatrix}
\begin{pmatrix} \hat{c}_{\omega} \\
                \hat{d}^{\dagger}_{\omega}\\
\end{pmatrix}
\end{align}
where $\tanh(r_\omega) = e^{- \pi \omega c/a}$ is defined in terms of the acceleration $a$ of the frame and the frequency $\omega$ of the field. This connects field operators in an accelerated frame on the left, to Unruh operators on the right whose vacuum is the Minkowski vacuum in the inertial frame.  \\

An evolution time $t$ for a given resonant mode ${\mathbf{k}_c}$ then corresponds to a frame transformation with acceleration  
\begin{equation}
    a = -\pi \omega c/\log(\tanh( |\Omega_{\mathbf{k}_c}| t/\hbar  ))
\end{equation}
and the excitations created in the quantum system during the time evolution thus correspond to the thermal occupation of modes observed in an accelerated frame, i.e. a process that quantum simulates Unruh radiation.\\

If we now take a closer look at the generated state after time evolution, which emerges from the vacuum state, it has the form,
\begin{equation}
    e^{i t h_{\mathbf{k}_c}} \lvert \mathrm{vac} \rangle = \frac{1}{\cosh(|\Omega_{\mathbf{k}_c}| t)} \sum_{n=0}^{\infty} \tanh^n(|\Omega_{\mathbf{k}_c}| t) \lvert n_{A,\mathbf{k}_c},n_{B,-\mathbf{k}_c}\rangle
\end{equation}
which is known in the literature as a thermo-field double state (TFD) \cite{TAKAHASHI1996,Zhu_PNAS_2020,Chapman_SciPost_2019}. This is a pure state of the form
\begin{equation}
    \frac{1}{\sqrt{Z}} \sum_n e^{- E n/(2 k_b T)} \lvert n \rangle_A \otimes \lvert n \rangle_B
\end{equation}
where the coefficients follow  a Boltzmann  distribution with a temperature set by the evolution time as 
\begin{equation}
    T= \frac{E}{2 k_B \log(\coth(|\Omega_{\mathbf{k}_c}| t)))}
\end{equation}
The Boltzmann-like   distribution generated  in  a  pure quantum state,  has the appeal that when each of the  modes of the TMS is  considered independently by tracing over the other,
\begin{equation}
    \rho_{\mathrm{reduced}} = \frac{1}{Z} \sum_n e^{- E n/(k_b T)} \lvert n \rangle \langle n \rvert
\end{equation}
the state  reduces to a thermal mixed state with an effective  temperature  $T$ \cite{TAKAHASHI1996}. These type of states have played a key role in the holographic correspondence relating a quantum-field theory to a gravitational theory in one higher dimension. In this correspondence, TFD states are dual to wormholes on the gravity side \cite{israel1976thermo,maldacena2003eternal} and enable teleportation (“transversable wormholes”) \cite{gao2017traversable,maldacena2017diving}.

\subsection{Comparison of open and closed boundary conditions}
\begin{figure}[htbp]
\centering
\includegraphics[width=\columnwidth]{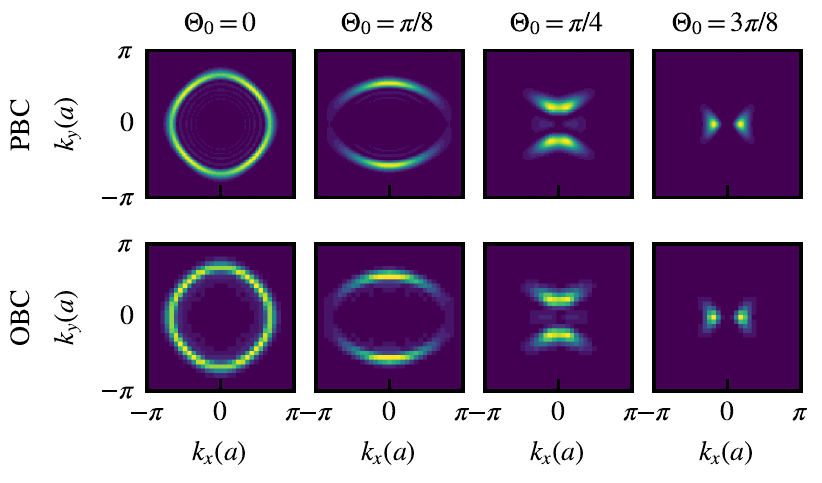}
\caption{Effect of the boundary conditions in the Bogoliubov analysis. We depict the quasi-momentum distribution of the created pairs, $N_k$, comparing periodic (top) and open (bottom) boundary conditions for a range of dipole orientations $\Theta_0$, at times such that $N_{\textrm{pair}}(t)= \sum_{\mathbf{k}} N_{\mathbf{k}}(t) = 0.1 N$. The results were obtained for a $33 \times 33$ bilayer with layer spacing $a_Z/a=2$ at unit filling.
\label{SI_fig:BG_OBC_PBC}
}
\end{figure}
The above mentioned procedure in real space for $f<1$ may be also employed for full filling, providing the time evolution in the presence of open-boundary conditions, rather than periodic boundary conditions, as implictly assumed in the analysis in quasi-momentum space. We consider the effects of boundary conditions on the momentum structure of the created pairs in finite systems within the Bogoliubov analysis in Fig.~\ref{SI_fig:BG_OBC_PBC}. We see that both periodic (top) and open boundaries (bottom) result in basically the same momentum structure across all dipole orientations. This demonstrates that the predicted phenomena should be accessible within the limitations on total particle numbers and lattice sizes available in experimental platforms.

\subsection{Comparison of momentum and real-space structure of correlations}
\begin{figure}
\includegraphics[width=\columnwidth]{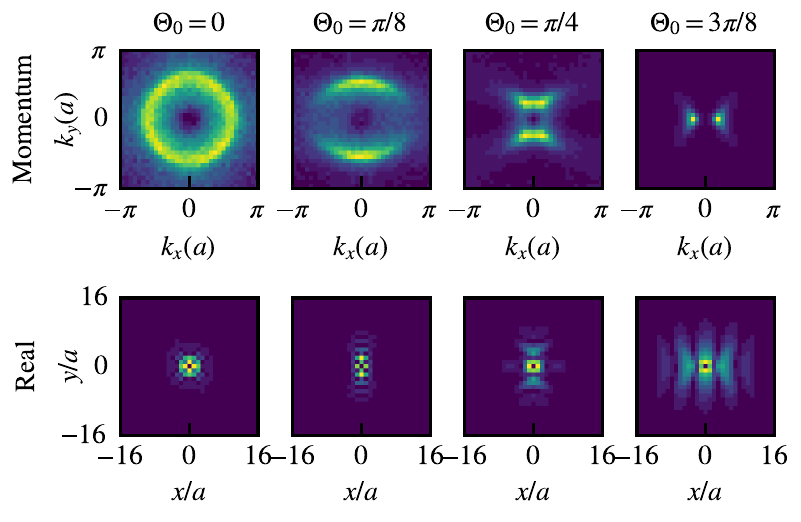}
\caption{Correlations in momentum and real space. Top panel spin-structure factor $S_{\mathbf{k}}^{+-}(t)$ (see \ref{eq_SI:spin_struct}), corresponding to momentum state population of pairs $N_k(t)$, compared to the real-space structure of spin-correlations $ |C^{+-}_{\mathbf{r}}(t)|$ (see \ref{eq_SI:spin_cor}). Results for a $33\times 33$ bilayer with a layer spacing of $a_Z/a=2$ and open boundary conditions at time $t$ such that $N_{\mathrm{pair}}(t) = N/10$. \label{fig_SI:real_space_structure}}
\end{figure}
In this section we provide results for the correlation structure in real space. While we mainly focus on the momentum structure of the correlations, as they directly map to the occupation of momentum modes and the Bogoliubov analysis, the real space correlations are what would be directly observed in an experiment with access to spatially resolved measurements.

Defining the spin-spin correlation function
\begin{equation}
 C^{+-}_{\mathbf{i}\mathbf{j}} = \left<\hat{s}_{\mathbf{i}}^{+} \hat{s}_{\mathbf{j}}^- \right>
 \label{eq_SI:spin_cor}
\end{equation}
the spin-structure factor $S_{\mathbf{k}}^{A(B),+-}$, which corresponds to the momentum mode occupation in the low excitation limit, is just
\begin{equation}
S_{\mathbf{k}}^{A(B),+-} =  \frac{1}{N}\sum_{\mathbf{i}\mathbf{j}\in A(B)} e^{ i \mathbf{k} (\mathbf{r}_{\mathbf{i}} -\mathbf{r}_{\mathbf{j}} )}  C^{+-}_{\mathbf{i}\mathbf{j}}
 \label{eq_SI:spin_struct}
\end{equation}
We emphasise that this directly corresponds to the connected correlation function for our initial state evolving under the $U(1)$-symmetric XXZ Hamiltonian which makes the one-point functions vanish identically at all times, i.e. $\langle \hat{s}_i^+\rangle =\langle \hat{s}_i^-\rangle=0$.

We compare these expressions directly in Fig.~\ref{fig_SI:real_space_structure} for a range of dipole orientations. The top panels shows the spin-structure factor $S^{+-}_{\mathbf{k}}$, and the bottom panels show the corresponding real-space correlation function $ C^{+-}_{\mathbf{r}}$ at a distance $\mathbf{r} = \mathbf{r_i}-\mathbf{r_j}$, both summed over the layers $A,B$. To make the structure of real-space correlations visible on top of the population growth, we only show them for $\mathbf{i} \neq \mathbf{j}$, e.g. set $ C^{+-}_{\mathbf{ii}}=0$.
These results highlight the intricate real-space structure of the correlations created during the pair-creation process. 
We note that up to boundary effects, the density of excitations itself is fully homogeneous throughout the dynamics, and the structure emerges within the inter-site off-diagonal correlations.

\appendix

\subsection{Extended results on time-dependence}
\begin{figure}
\includegraphics[width=\columnwidth]{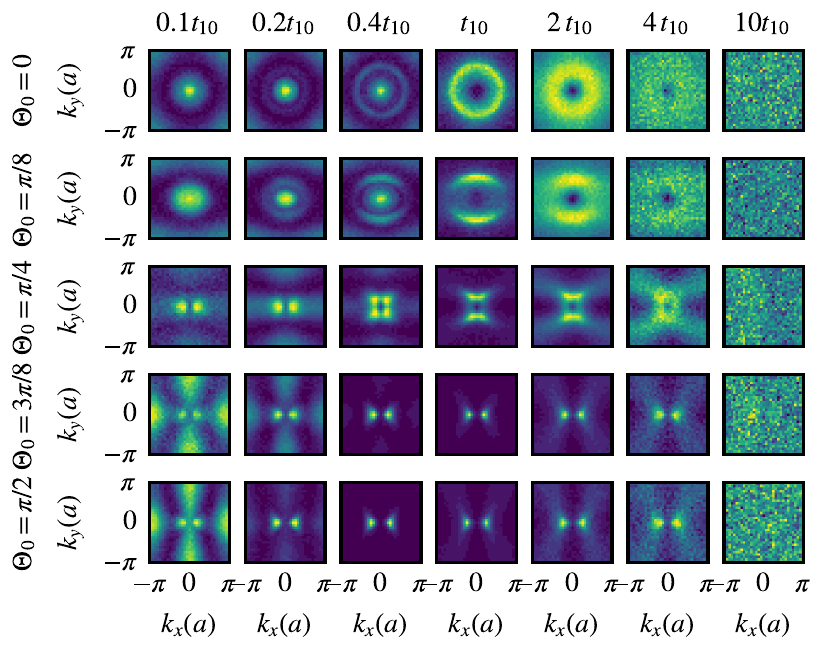}
\caption{Extended DTWA results on time evolution. Momentum state population of pairs $N_k(t)$ within DTWA showing the different regimes of dynamics for a range of dipole orientations $\Theta_0$ at times in terms of $t_{10}$ at which $N_{\mathrm{pair}}(t_{10})= 0.1 N$. The left most panel shows the early time regime before  exponential growth has taken over. The second and third panel show the build up of the expected momentum structure. The central panel shows the fully built-up expected momentum structure. The last three panels show the subsequent thermalisation as scattering between momentum modes occurs.  Results for a $33\times 33$ bilayer with a layer spacing of $a_Z/a=2$ and open boundary conditions.\label{fig_SI:time_dependence}}
\end{figure}
We provide extended results for the time-dependence of the momentum structure of the created pairs obtained within DTWA in Fig.~\ref{fig_SI:time_dependence}. This provides both the full range of dipole orientations (in contrast to the single case of $\Theta_0=3
\pi/8$ in the main text), as well as additional times during the build-up of correlations, as well as during the late time thermalisation state. The qualitative picture remains the same for all dipole orientations, in that at very early times, the dynamics of stable modes can dominate over the exponentially growing unstable modes, which establish the expected momentum structure at intermediate times, before scattering between modes leads to thermalisation and a homogeneous background at late times.

\subsection{Results at finite filling}
\begin{figure}
\includegraphics[width=\columnwidth]{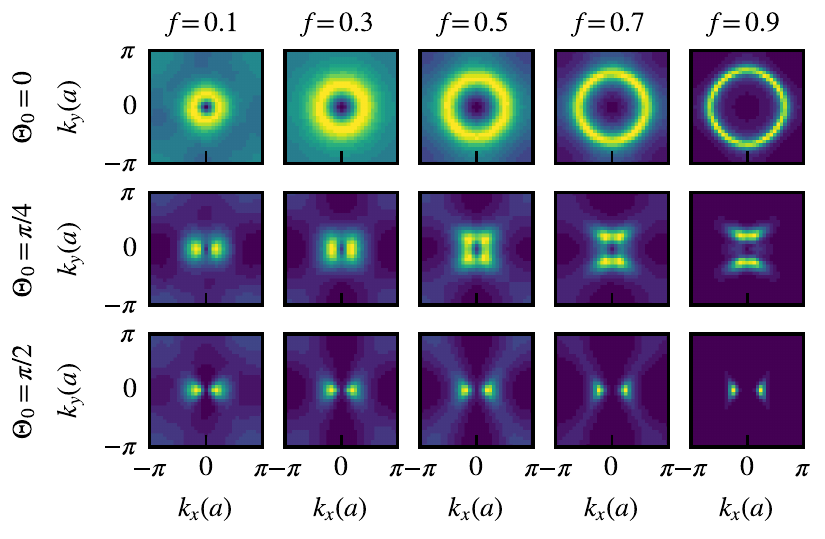}
\caption{Bogoliubov results at finite filling. Momentum state population $N_k$ for a range of $\Theta_0$ and filling fraction $f$ at times such that $N_{\textrm{pair}}(t)= 0.1 f N$. Results for a $33\times 33$ bilayer with a layer spacing of $a_Z/a=2$ and open boundary conditions.\label{fig_SI:BDG_momentum_structure_disorder}}
\end{figure} 
\begin{figure}
\includegraphics[width=\columnwidth]{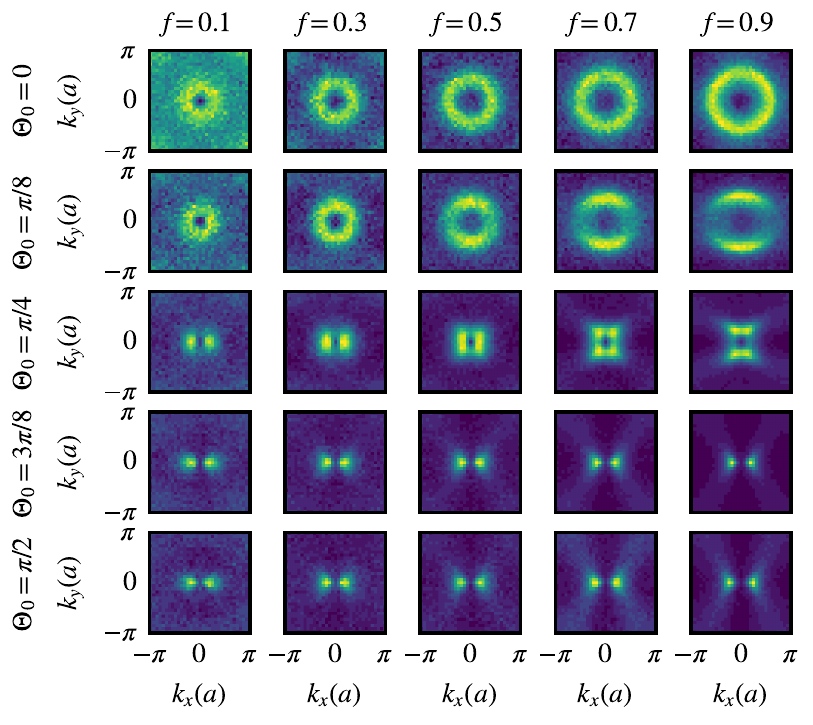}
\caption{Extended DTWA results at finite filling. Momentum structure of created pairs $N_{\mathbf{k}}(t)$ at times $t$ such that the total number of pairs $N_{pair}(t) =  0.1 f N$. Results for a range of $\Theta_0$ and filling fraction $f$ for a $33\times 33$ bilayer with a layer spacing of $a_Z/a=2$  and open boundary conditions  \label{SI_fig:momentum_structure_disorder}}
\end{figure} 
Figure~\ref{fig_SI:BDG_momentum_structure_disorder} shows the quasi-momentum distribution of the created pairs for an imperfect filling within the Bogoliubov analysis, following the procedure discussed above. These results should be compared with the results shown in Fig.3(b) of the main text, as well as with the results covering an expanded set of dipole orientations in Fig.~\ref{SI_fig:momentum_structure_disorder}. 
The DTWA results are averaged over $10000$ realisations of the lattice occupations, whereas the Bogoliubov results average over $200$ realisations.

Across all dipole orientations and filling fractions we observe again a very good agreement between the spin dynamics obtained with the DTWA and the Bogoliubov predictions. In particular, both show the shrinking of the structures in momentum space as the filling fraction is lowered. Intuitively, large momentum modes would be expected to be more strongly affected by the introduction of local disorder, whereas small momentum modes would be expected to be more resilient, which seems to be the case here. 

Importantly, the dynamics remains qualitatively unaffected by the imperfect filling, being still characterized by the exponential growth of characteristic patterns in quasi-momentum space, that depend on the dipole orientation. This robustness against imperfect filling is particularly relevant, since it makes feasible the observation of the effect in current experimental platforms.

\end{document}